\documentclass[showpacs,amssymb,twocolumn,aps,nofootinbib]{revtex4}
\usepackage{amsmath}
\usepackage{amstext}
\usepackage{amsopn}
\usepackage{amsfonts}
\usepackage{amssymb}
\usepackage{comment}
\usepackage{accents}
\usepackage{bbm}
\usepackage{empheq}
\usepackage{graphicx}
\usepackage{epsf}
\usepackage[sort&compress]{natbib}
\usepackage{float}
\usepackage{subfigure}
\usepackage{graphics}
\usepackage{xcolor}
\usepackage[latin1]{inputenc}

\allowdisplaybreaks[3]

\begin{document}
\title{Fermion propagator in an external potential and generalized Airy functions}	

\author{A. L. M. Britto$^{a}$, Ashok K. Das$^{b,c}$ and J. Frenkel$^{a}$}
\affiliation{$^{a}$ Instituto de F\'{i}sica, Universidade de S\~{a}o Paulo, 05508-090, S\~{a}o Paulo, SP, Brazil}
\affiliation{$^b$ Department of Physics and Astronomy, University of Rochester, Rochester, NY 14627-0171, USA}
\affiliation{$^c$ Institute of Physics, Sachivalaya Marg, Bhubaneswar 751005, India}


\begin{abstract}
We study the behavior of the fermion propagator in an external time dependent potential in 0+1 dimension. We show that,  when the potential has upto quadratic terms in time, the propagator can be expressed in terms of generalized Airy functions (or standard Airy functions depending on the exact time dependence). We study various properties of these new generalized functions which reduce to the standard Airy functions in a particular  limit. 
\end{abstract}
\maketitle

\section{Introduction}

It is well known that a linear potential between quarks in QCD (Quantum Chromo Dynamics) leads to confinement of quarks. The time independent radial equation of a three dimensional Schr\"{o}dinger equation (for $\ell=0$) or that of a one dimensional equation, with a linear potential, can be written as (with different boundary conditions for the two cases)
\begin{equation}
\left[\frac{d^{2}}{dr^{2}} +(E_{n} - a_{1} r)\right] u(r) = 0,
\end{equation}
where $a_{1}$ is a real constant denoting the strength of the linear potential. With a suitable change of variables, this equation can be rewritten as
\begin{equation}
\frac{d^{2}u (\bar{r})}{d \bar{r}^{2}} - \bar{r} u (\bar{r}) = 0,
\end{equation}
which is the Airy equation \cite{AS,dasquantum}. This is how the Airy functions enter into quantum mechanics and to the physical question of quark confinement. It also shows up in many other branches of physics.

Phenomenologically, the non-relativistic (static) potential used in the study of quarkonium in QCD is conventionally taken to be of the form \cite{bali}
\begin{equation}
V(r) = - \frac{\alpha}{r} + a_{1} r,
\end{equation}
where $\alpha$ is the fine structure constant. The first term, on the right, represents the standard tree level Coulmb potential between a quark and an antiquark whereas the second linear term represents the confining potential. If we ignore the Coulomb potential (or consider the large distance behavior), the exact solutions of the Schr\"{o}dinger equation will lead to Airy functions. More recently it has been shown  \cite{sumino} that higher order calculations, taking into account the cancellation of leading order renormalons and other nonperturbative effects, lead to a generalization of the potential of the form 
\begin{equation}
V (r) = -\frac{\alpha}{r} + a_{0} + a_{1} r + a_{2} r^{2},\label{qcdpotential}
\end{equation}
where the constants $a_{0}, a_{2}$ are real constants and have nonzero contributions beginning at two loops. Even if we ignore the Coulomb potential, the exact solutions in this case would not correspond to Airy functions in general.

All of this discussion is within the context of the second order time independent Schr\"{o}dinger equation. In this paper we study the solutions for a one dimensional fermion interacting with a time dependent external potential. More specifically, we study the behavior of the fermion propagator in an external potential. Motivated by the QCD results \eqref{qcdpotential}, we choose the form of the external potential to be
\begin{equation}
A(t) = a_{0} + a_{1} t + a_{2} t^{2},\label{externalpotential}
\end{equation}
where $a_{0},a_{1}, a_{2}$ are real constants. We show that in this first order system of equation (to be discussed below), depending on the values of the constant $a_{1}$, the propagator is described either by Airy functions or by generalized Airy functions. In particular, if $a_{1}=0$ (absence of a linear term in $t$ in the potential), the propagator is described in terms of a linear combination of the Airy and the Scorer functions \cite{AS} which should be contrasted with the time independent (second order) case. On the other hand, if $a_{1}\neq 0$, the propagator is described in terms of generalized Airy functions which reduce to Airy functions in an appropriate limit. These generalized Airy functions are new and we study their properties systematically. We note here that, in view of the importance of the Airy functions in physics as well as in mathematics, various other generalizations of these functions have also been studied previously \cite{drazin,swanson,Chin,Janson,Kamimoto}.

The paper is organized as follows. In section {\bf II}, we study the fermion propagator systematically and point out the physical meaning of the constants in \eqref{externalpotential}. We show how generalized Airy functions arise when $a_{1}\neq 0$ and how these generalized functions reduce to the conventional Airy functions when $a_{1}=0$. In section {\bf III}, we discuss, in detail, various properties of these generalized functions and how they reduce to the standard results in an appropriate limit. We present a brief conclusion in section {\bf IV}. In the appendix, we derive the propagator by solving the differential equation in the Fourier transformed space and compare it with the coordinate space calculations. 

\section{Fermion propagator}
In $0+1$ dimension, the fermion propagator in a time dependent external field, satisfies the first order differential equation
\begin{equation}
(i\partial_t -m-gA(t))S(t)=i\delta(t),\label{propagatoreqn}
\end{equation}
where $m$ is the mass of the fermion, $g$ is the coupling constant and $A(t)$ is a time dependent real external potential which can, in principle, be the gauge potential (in that case $g=e$). In $0+1$ dimension, the Feynman propagator coincides with the retarded propagator. Therefore, if we  assume that the potential switches on at $t=0$, the unique (Feynman/retarded) solution of equation \eqref{propagatoreqn} can be written as 
\begin{equation}
S(t)= \theta(t)\,e^{-imt}\, e^{-ig\int_0^ t dt'A(t')},\label{propagatorgensoln}
\end{equation}
where $\theta(t)$ denotes the step function.

In principle, the second exponential factor in \eqref{propagatorgensoln} with time integration should involve a time-ordered product. However, since $A(t)$ denotes a classical external potential, time ordering is not relevant. Furthermore, we note that, in $0+1$ dimension, the fermion propagator coincides with the time evolution operator for the time dependent Hermitian Hamiltonian $H(t)=m + gA(t)$, namely, we can also write $S(t) = U(t) = \theta(t) e^{-i\int_{0}^{t} dt'\,H(t')}$. It follows from this that formally (time evolution is unitary)
\begin{equation}
S(t) S^{\dagger}(t) = \mathbbm{1} = S^{\dagger}(t) S(t),\label{unitary}
\end{equation}
namely, the fermion propagator, in coordinate space, is formally unitary.

The propagator, in the Fourier transformed space, can be obtained to be
\begin{align}
S(p) & = \int_{-\infty}^{\infty} dt\,e^{ipt}\,S(t)\nonumber\\
& = \int_{0}^{\infty} dt\, e^{i(p-m)t}\, e^{-ig\int_{0}^{t} dt'\, A(t')},\label{S(p)}
\end{align}
which satisfies the differential equation (in the Fourier transformed space)
\begin{equation}
\left(p-m -g A(-i\frac{d}{dp})\right) S(p) = i.\label{S(p)eqn}
\end{equation}

If $g=0$ (or $A(t)=0$), namely, if there is no interaction with an external potential, \eqref{S(p)} (or \eqref{S(p)eqn}) yields the free fermion propagator
\begin{equation}
S_{0}(p) = \frac{i}{p-m},\label{S_{0}(p)}
\end{equation}
where the Feynman $i\epsilon$ prescription is understood and is necessary for the convergence of the integral in \eqref{S(p)}. (Throughout the paper, we will suppress the $i\epsilon$ term for simplicity.)

The complete fermion propagator, in the Fourier transformed (energy) space, for the theory can be determined either by evaluating the integral in \eqref{S(p)} or by solving the differential equation in \eqref{S(p)eqn}. In this section, we will determine $S(p)$ by evaluating the integral in \eqref{S(p)}. In the appendix, we will show that this result coincides with the one obtained by solving  \eqref{S(p)eqn}. Of course, the propagator cannot be explicitly determined in a closed form  for an arbitrary potential. Therefore, motivated by the studies in QCD, we will choose the potential to be of the form (see \eqref{externalpotential})
\begin{equation}
A(t) = a_{0} + a_{1} t + a_{2} t^{2},\label{A(t)}
\end{equation}
where $a_{0}, a_{1}, a_{2}$ are real constants. 

We note that, in this case, we can determine a superpotential $W(t)$ such that
\begin{align}
A(t) & = \frac{dW(t)}{dt},\nonumber\\
W(t) & = C + a_{0} t + \frac{a_{1}}{2} t^{2} + \frac{a_{2}}{3} t^{3},\label{W(t)}
\end{align} 
where $C$ is a constant of integration. In terms of the superpotential, we can rewrite \eqref{S(p)} as
\begin{equation}
S(p) = \int_{0}^{\infty} dt\, e^{i(p-m)t -ig(W(t)-W(0))}.\label{S(p)'}
\end{equation}
We note here for future use that
\begin{equation}
W(t)-W(0) = a_{0} t + \frac{a_{1}}{2} t^{2} + \frac{a_{2}}{3} t^{3},\label{exponent}
\end{equation}
where the arbitrary constant has canceled out.

In the next three subsections, we evaluate $S(p)$ from \eqref{S(p)'} and \eqref{exponent} systematically and comment on the physical meaning of the three terms in the potential. In the process we will discover how generalized Airy functions show up in this study.

\subsection{$A(t)=a_{0}$}

In this case, we have $a_{1}=a_{2}=0$ so that \eqref{exponent} yields
$$
W (t) - W(0) = a_{0}t,
$$
and \eqref{S(p)'} leads to
\begin{equation}
S(p) = \frac{i}{p-m-ga_{0}} = \frac{i}{p-M},\label{a_{0}}
\end{equation}
where we have identified 
\begin{equation}
M= m+ga_{0}.\label{M}
\end{equation}
Thus, for a constant external potential, the fermion propagator continues to be a free propagator, but with a shifted mass $M$.

\subsection{$A(t) = a_{0} + a_{1}t$}

In this case, we have $a_{2}=0$ so that \eqref{exponent} yields
$$
W(t) - W(0) = a_{0}t + \frac{a_{1}}{2} t^{2},
$$
and \eqref{S(p)'} leads to
\begin{equation}
S(p) = \int_{0}^{\infty} dt\, e^{i(p-M)t - \frac{iga_{1}}{2} t^{2}}.
\end{equation}
Completing the square in the exponent in the integrand (and redefining variables) we obtain
\begin{align}
S(p) & = \sqrt{\frac{2}{iga_{1}}}\,e^{\frac{i(p-M)^{2}}{2ga_{1}}}\notag\\
& \times\left[\int_{0}^{\infty} dt\, e^{-t^{2}} - \int_{0}^{\sqrt{\frac{iga_{1}}{2}} \frac{M-p}{ga_{1}}} dt\,e^{-t^{2}}\right]\notag\\
& = \sqrt{\frac{\pi}{2iga_{1}}}e^{\frac{i(p-M)^{2}}{2ga_{1}}}\left[1 - \Phi (\sqrt{\frac{i}{2ga_{1}}} (M-p))\right],\label{a_{1}}
\end{align}
where
\begin{equation}
\Phi (x) = \frac{2}{\sqrt{\pi}}\int_{0}^{x} dt\, e^{-t^{2}}\label{prob}
\end{equation}
denotes the probability function or the error function.

For small values of $ga_{1}$, the argument of the probability function in \eqref{a_{1}} becomes large and the asymptotic expansion \cite{GR}
\begin{equation}
\Phi (x)\xrightarrow{x\rightarrow \infty} 1 - \frac{1}{\sqrt{\pi}}\,\frac{e^{-x^{2}}}{x} + \frac{1}{2\sqrt{\pi}}\,\frac{e^{-x^{2}}}{x^{3}}+\cdots,\label{probexpansion}
\end{equation}
leads to
\begin{align}
S(p) & \rightarrow \frac{i}{p-M}\left(1 + \frac{iga_{1}}{(p-M)^{2}}+\cdots\right)\notag\\
& \simeq \frac{i}{p-M}\, \frac{1}{1-\frac{iga_{1}}{(p-M)^{2}}}\nonumber\\
& = \frac{i}{p-M - \frac{iga_{1}}{p-M}},\label{S_{1}(p)}
\end{align}
where we have assumed $ga_{1}\ll (p-M)^{2}$. Therefore, to linear order in $ga_{1}$, the pole of the propagator can be determined to be at (we assume $ga_{1} > 0$)
\begin{equation}
p = \mu = M \pm \sqrt{iga_{1}} = M \pm (1+i)\,\sqrt{\frac{ga_{1}}{2}}.\label{sqrt}
\end{equation}

While both the signs of the square root in \eqref{sqrt} are mathematically allowed, we choose the negative sign that leads to a retarded propagator (which is what we are studying here). (The positive sign would lead to an advanced propagator.) In this case, we can write the pole of the propagator to be at
\begin{equation}
p = \mu = \mu_{R} - i\mu_{I},\label{pole}
\end{equation}
where
\begin{equation}
\mu_{R} = M - \sqrt{\frac{ga_{1}}{2}},\qquad \mu_{I} = \sqrt{\frac{ga_{1}}{2}}.
\end{equation}
The presence of an imaginary part in the pole denotes a Breit-Wigner form of the solution which represents a decay in the amplitude because of (scattering) interaction with the external potential.

\subsection{$A(t) = a_{0} + a_{1}t + a_{2}t^{2}$}

In this case, we have (see \eqref{exponent})
$$
W(t) - W(0) = a_{0} t + \frac{a_{1}}{2} t^{2} + \frac{a_{2}}{3} t^{3},
$$
and \eqref{S(p)'} leads to
\begin{equation}
S(p) = \int_{0}^{\infty} dt\, e^{i(p-M)t - \frac{iga_{2}}{3} (t^{3} + \frac{3a_{1}}{2a_{2}} t^{2})}.
\end{equation}
Completing the cube in the exponent of the integrand (and redefining variables), this leads to
\begin{align}
S(p) & = e^{-\frac{ia_{1}}{2a_{2}} (p-M + \frac{ga_{1}^{2}}{6a_{2}})}\notag\\
&\quad \times \int_{\frac{a_{1}}{2a_{2}}}^{\infty} dt\, e^{i (p-M + \frac{ga_{1}^{2}}{4a_{2}})t - \frac{iga_{2}}{3} t^{3}}.\label{a_{2}int}
\end{align} 

If we assume that $ga_{2} > 0$ (we will comment on the case $ga_{2} < 0$ later) and define
\begin{equation}
u = (ga_{2})^{\frac{1}{3}}\, t,\ q = \frac{ga_{1}}{2 (ga_{2})^{\frac{2}{3}}},\ x = -\frac{(p-M+ \frac{ga_{1}^{2}}{4a_{2}})}{(ga_{2})^{\frac{1}{3}}},\label{defn}
\end{equation}
so that
\begin{equation}
x+q^{2} = \frac{p-M}{(ga_{2})^{\frac{1}{3}}},\label{defn1}
\end{equation}
we can write the propagator in \eqref{a_{2}int} as
\begin{equation}
S(p) = \frac{\pi\, e^{i(qx+\frac{q^{3}}{3})}}{(ga_{2})^{\frac{1}{3}}}\, F_{q} (x),\label{prop}
\end{equation}
where we have identified
\begin{equation}
F_{q} (x) = \frac{1}{\pi} \int_{q}^{\infty} du\, e^{-i (ux + \frac{u^{3}}{3})}.\label{F}
\end{equation}

The study of the propagator is, therefore, equivalent to the study of the function $F_{q}(x)$ which we will do in the rest of the paper. We note that the integral in \eqref{F} is reminiscent of the Airy function. In fact, for real $q$, if we define
\begin{align}
Ai_{q} (x) & = \frac{1}{\pi} \int_{q}^{\infty} du\,\cos (ux + \frac{u^{3}}{3}) = \text{Re}\, F_{q}(x),\notag\\
Gi_{q} (x) & = \frac{1}{\pi} \int_{q}^{\infty} du\, \sin (ux + \frac{u^{3}}{3}) = -\text{Im}\, F_{q}(x),\label{generalizedairy}
\end{align}
we can write
\begin{equation}
F_{q} (x) = Ai_{q} (x) - i Gi_{q} (x).\label{F'}
\end{equation}
We call $Ai_{q} (x)$ and $Gi_{q} (x)$ the generalized Airy and Scorer functions, since in the limit $q=0$, they reduce to the Airy and the Scorer functions (we note here parenthetically that $q=0$ corresponds to $a_{1}=0$ for $g\neq 0$) \cite{AS}
\begin{align}
Ai_{0} (x) & = \frac{1}{\pi} \int_{0}^{\infty} du\,\cos (ux + \frac{u^{3}}{3}) = Ai (x),\notag\\
Gi_{0} (x) & = \frac{1}{\pi} \int_{0}^{\infty} du\,\sin(ux + \frac{u^{3}}{3}) = Gi (x).\label{airy}
\end{align}
This also allows us to identify
\begin{equation}
F(x) = F_{0}(x) = Ai(x) - iGi(x).\label{Fzero}
\end{equation}

The Airy function can also be identified with the propagator alternatively as
\begin{align}
Ai (x) & = \frac{1}{\pi}\int_{0}^{\infty} du\,\cos(ux+\frac{u^{3}}{3})\notag\\
& = \frac{1}{2\pi} \int_{-\infty}^{\infty} du\, e^{-i (ux+\frac{u^{3}}{3})} = \frac{1}{2}\,F_{-\infty} (x).
\end{align}
This shows how intricately the generalized Airy functions are connected with the fermionic propagator in an external (quadratic) potential. Namely, in a general quadratic potential without a linear term ($a_{1}=0, q=0$), the propagator is given as a combination of the standard Airy and Scorer functions. However, if the linear term is present in the potential, the Airy and Scorer functions modify to what we call the generalized Airy functions. We also note here that if $ga_{2} < 0$, this would simply correspond to replacing
\begin{equation}
F_{q} (x) \rightarrow (F_{-q} (-x))^{*}.
\end{equation}

In what follows, we will study various properties of these generalized Airy and Scorer functions.

\section{Properties of the generalized Airy functions}

In this section, we will discuss various properties of the generalized Airy functions.

\subsection{Equations satisfied by the generalized functions}

Let us recall that the Airy function $Ai (x)$ and the Bairy function $Bi (x)$ satisfy the homogeneous differential equation \cite{AS}
\begin{equation}
\frac{d^{2} y(x)}{dx^{2}} - x y(x) = 0. \label{airyeqn}
\end{equation}
They are the two independent solutions of \eqref{airyeqn} with the Wronskian
\begin{equation}
W\{Ai(x), Bi(x)\} = \frac{1}{\pi}.\label{wronskian}
\end{equation}
Similarly, the Scorer function $Gi (x)$ (as well as $-Hi (x)$) satisfy the inhomogeneous equation
\begin{equation}
\frac{d^{2} y(x)}{dx^{2}} - x y(x) =  - \frac{1}{\pi}.\label{scorereqn}
\end{equation}
The Scorer function can be expressed as a linear combination of $Ai (x)$ and $Bi (x)$ as
\begin{equation}
Gi (x) = \frac{1}{3} Bi (x) + \int_{0}^{x} dt\,(Ai(x) Bi(t) - Ai(t) Bi(x)).\label{linearcombination}
\end{equation}

In a similar manner starting from the definitions in \eqref{generalizedairy}, one can show that the generalized Airy and  Scorer functions satisfy the ($x$ dependent) inhomogeneous equations
\begin{align}
\frac{\partial^{2} Ai_{q}(x)}{\partial x^{2}} - x Ai_{q}(x) & = \frac{1}{\pi}\,\sin (qx + \frac{q^{3}}{3}),\notag\\
\frac{\partial^{2} Gi_{q} (x)}{\partial x^{2}} - x Gi_{q} (x) & = -\frac{1}{\pi}\, \cos (qx + \frac{q^{3}}{3}).\label{generalizedeqns}
\end{align}
It is clear that for $q=0$, these equations reduce to the Airy equation \eqref{airyeqn} and the Scorer equation \eqref{scorereqn} respectively. Furthermore, it follows from \eqref{generalizedeqns} as well as the identification in \eqref{F'} (or directly from \eqref{F}) that $F_{q}(x)$ satisfies the ($x$ dependent) inhomogeneous equation
\begin{equation}
\frac{\partial^{2} F_{q}(x)}{dx^{2}} - xF_{q} (x) = \frac{i}{\pi}\, e^{-i(qx + \frac{q^{3}}{3})} = -i\, \frac{\partial F_{q} (x)}{\partial q}.\label{Feqn}
\end{equation}

The generalized functions $Ai_{q}(x), Gi_{q}(x)$ as well as $F_{q}(x)$ can also be expressed as linear combinations of the independent functions $Ai(x), Bi(x)$. For example, we can write
\begin{align}
F_{q}(x) & = a(q) Ai (x) + b(q) Bi(x)\notag\\
&\quad - i e^{-\frac{iq^{3}}{3}}\!\!\!\int_{-q^{2}}^{x} \!\!dt\,(Ai(x)Bi(t) - Bi(x)Ai(t))e^{-iqz},\label{linearsuperposition}
\end{align}
where the coefficients $a(q),b(q)$ are given by
\begin{align}
a(q) & = -\pi (Bi (-q^{2})F'_{q}(-q^{2}) - Bi'(-q^{2})F_{q}(-q^{2})),\notag\\
b(q) & = \pi (Ai(-q^{2})F'_{q}(-q^{2}) - Ai'(-q^{2})F_{q}(-q^{2})),\label{constants}
\end{align}
with
\begin{align}
F_{q}(-q^{2}) & = \left. F_{q}(x)\right|_{x=-q^{2}} = \frac{e^{\frac{2iq^{3}}{3}}}{\pi}\int_{0}^{\infty} du\,e^{-i(u^{2}q + \frac{u^{3}}{3})},\notag\\
F'_{q}(-q^{2}) & = \left.\frac{\partial F_{q}(x)}{\partial x}\right|_{x=-q^{2}}\notag\\
 & = \frac{-ie^{\frac{2iq^{3}}{3}}}{\pi}\int_{0}^{\infty} du\,(q+u)\,e^{-i(u^{2}q+ \frac{u^{3}}{3})}.\label{FF'}
\end{align}
The value of the constants $a(0), b(0)$ can be easily calculated from \eqref{constants} using the definitions \eqref{Fzero}, \eqref{FF'}, the Wronskian in \eqref{wronskian} as well as \eqref{linearcombination} and leads to
\begin{equation}
a(0) = 1,\quad b(0) = - \frac{i}{3}.
\end{equation}
With this, it follows easily that \eqref{linearsuperposition} reduces to \eqref{Fzero} when $q=0$ (using \eqref{linearcombination}). Recalling that (see, for example, \eqref{generalizedairy})
\begin{align}
Ai_{q} (x) & = \text{Re}\, F_{q}(x),\notag\\
Gi_{q} (x) & = - \text{Im}\, F_{q} (x),
\end{align}
we can also obtain the expressions for $Ai_{q}(x), Gi_{q}(x)$ in terms of the two independent solutions of \eqref{airyeqn} by taking the real and the imaginary parts of \eqref{linearsuperposition}. 

\subsection{Fourier transformation}

From the definition \eqref{F}
$$
F_{q}(x) = \frac{1}{\pi} \int_{q}^{\infty} du\,e^{-i(ux + \frac{u^{3}}{3})},
$$
we note that the Fourier transform can be defined as
\begin{align}
\widetilde{F}_{q} (k) & = \int_{-\infty}^{\infty} dx\,e^{ikx} F_{q}(x)\notag\\
 & = \frac{1}{\pi}\int_{-\infty}^{\infty} dx\int_{q}^{\infty} du\,e^{-\frac{iu^{3}}{3}}\,e^{i(k-u)x}.\label{FT0}
 \end{align}
Interchanging the order of integration, this can be evaluated in a straightforward manner to give a simple form
 \begin{equation}
 \widetilde{F}_{q}(k) = 2\theta (k-q)\,e^{-\frac{ik^{3}}{3}}.\label{FT}
 \end{equation}
It is worth pointing out here that $k$ is not the energy, rather it is the conjugate variable to $x$ (defined in \eqref{defn}). 
 
In a similar manner, starting from the definitions in \eqref{generalizedairy}, we can derive 
 \begin{align}
 \widetilde{Ai}_{q}(k) & = (\theta(k-q) + \theta(-k-q))\,e^{-\frac{ik^{3}}{3}},\notag\\
 \widetilde{Gi}_{q}(k) & = i (\theta(k-q) - \theta(-k-q))\, e^{-\frac{ik^{3}}{3}}.\label{FT1}
 \end{align}
In the limit $q\rightarrow 0$, these lead to the well known results
 \begin{equation}
 \widetilde{Ai}(k) = e^{-\frac{ik^{3}}{3}},\quad \widetilde{Gi}(k) = i\text{sgn}(k)\,e^{-\frac{ik^{3}}{3}},
 \end{equation}
 where ${\rm sgn} (k)$ denotes the sign function.

\subsection{Graphical behavior}
\begin{figure}[H]
\subfigure{}\includegraphics[scale=0.7]{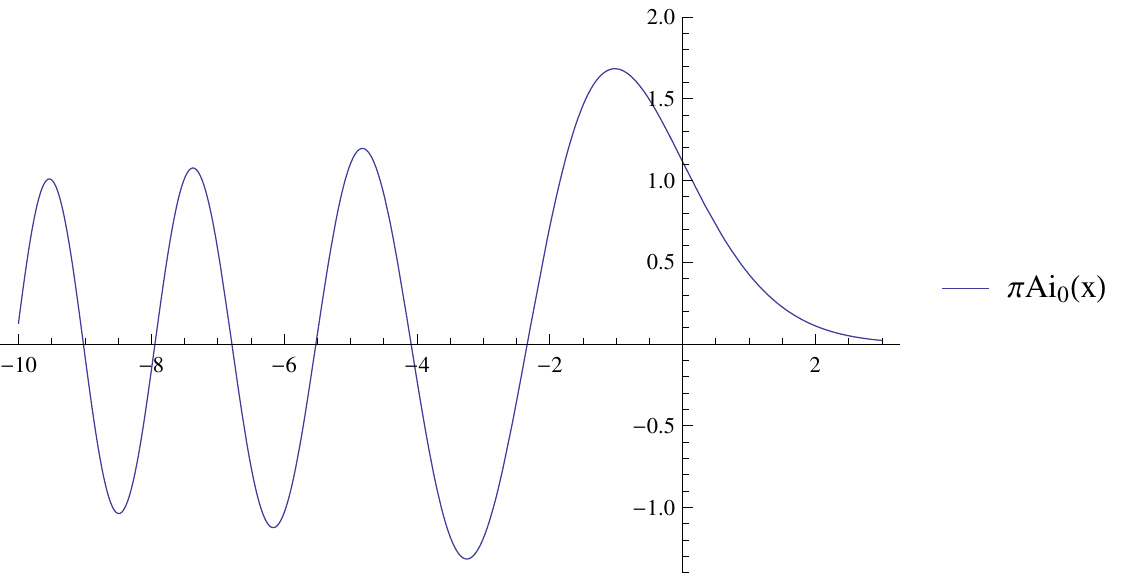}
\qquad
\subfigure{}\includegraphics[scale=0.7]{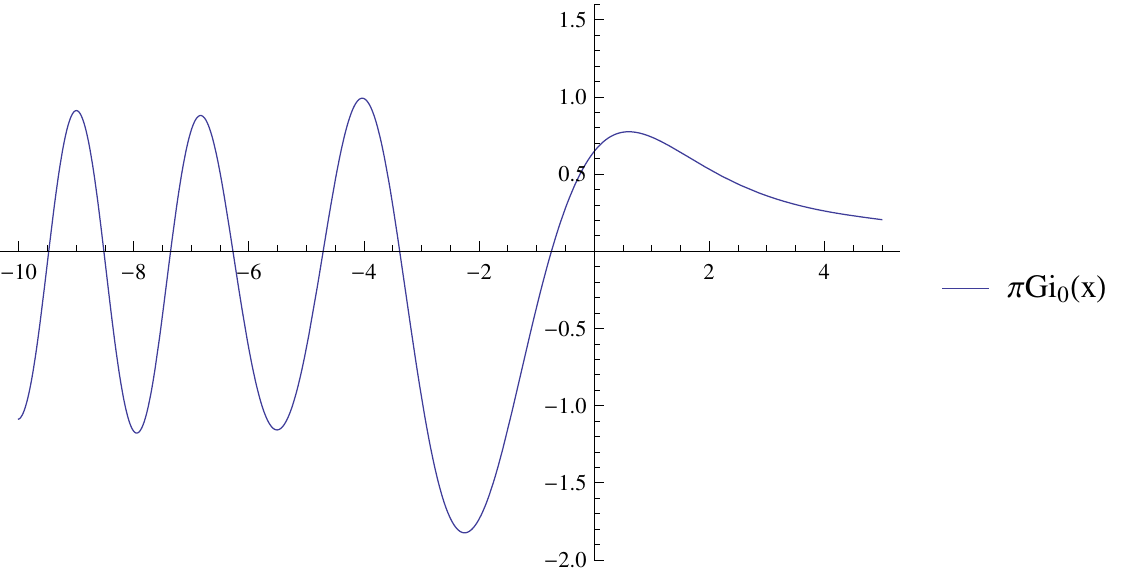}
\caption{Behavior of $Ai(x)=Ai_0(x)$ and $Gi(x)=Gi_0(x)$.}
\label{f1}
\end{figure}

The behavior of the Airy function, $Ai(x)=Ai_{0}(x)$, as well as the Scorer function, $Gi(x)=Gi_{0}(x)$, are well known and have the forms shown in Fig. \ref{f1}. The functions are well behaved at the origin as well as when $x\rightarrow \infty$ and are oscillatory for $x <0$. In fact, both the functions vanish as $x\rightarrow \infty$.

\begin{figure}[h]
\subfigure{}\includegraphics[scale=0.7]{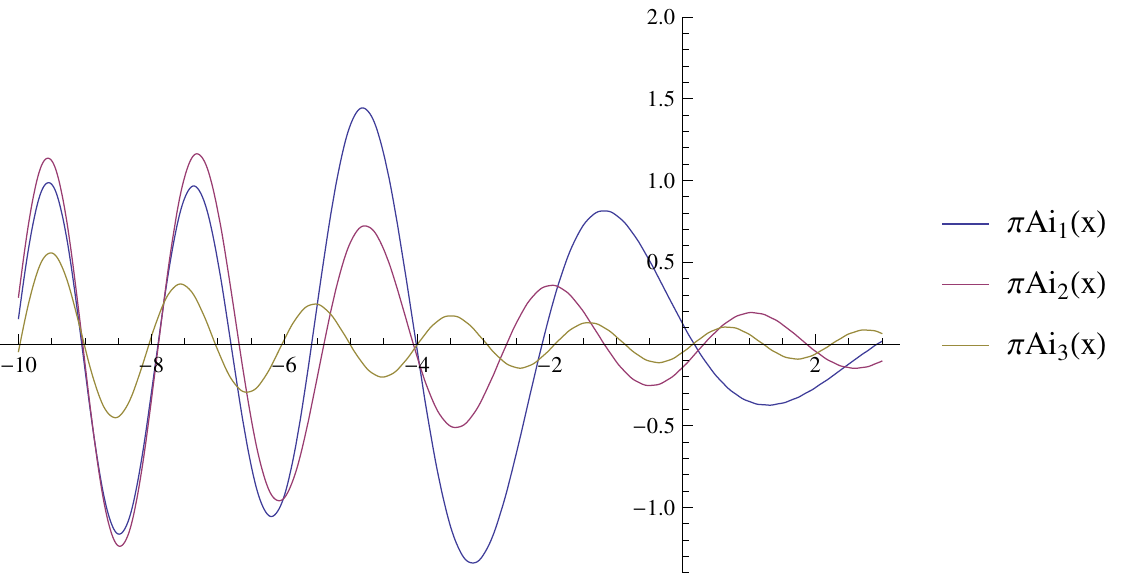}
\subfigure{}\includegraphics[scale=0.7]{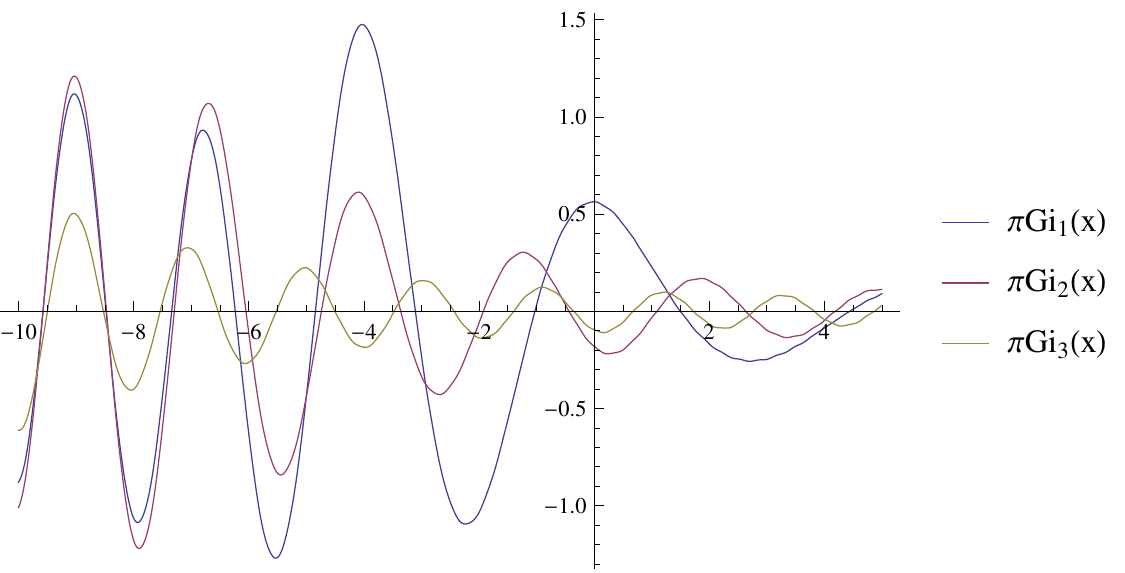}
\caption{Behavior of $Ai_q(x)$ and $Gi_q(x)$ for positive integer $q$.}
\label{f2}
\end{figure}

\begin{figure}[h]
\subfigure{}\includegraphics[scale=0.7]{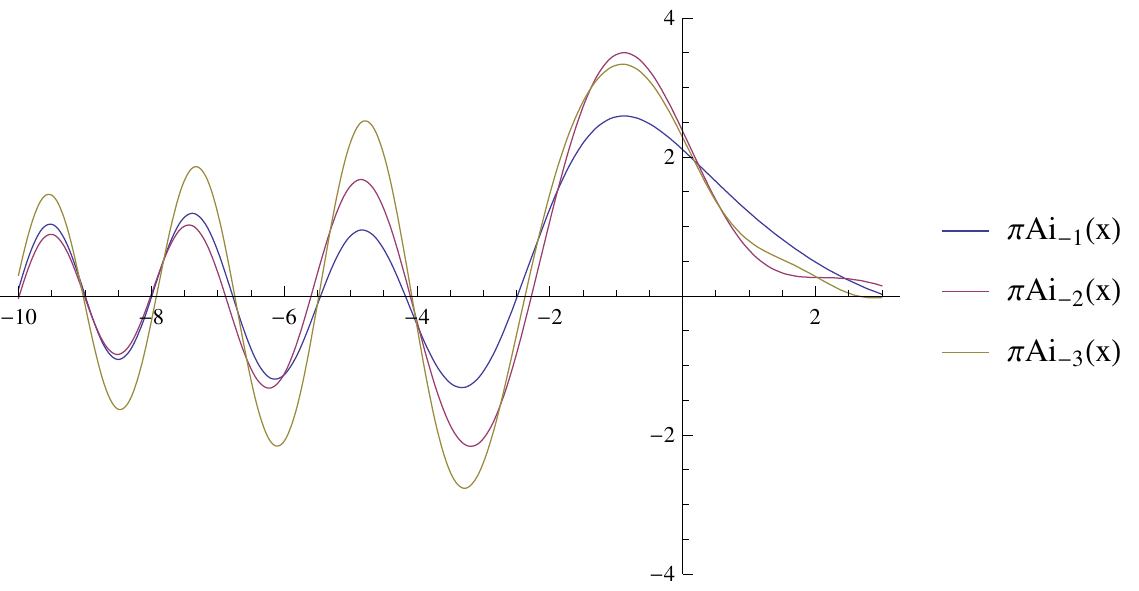}
\subfigure{}\includegraphics[scale=0.7]{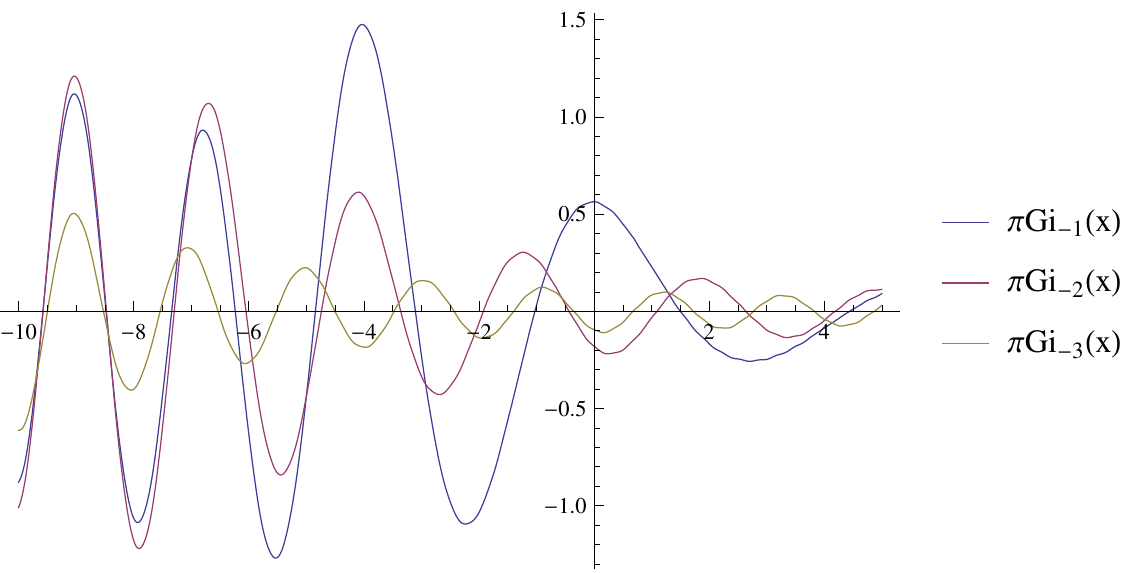}
\caption{Behavior of $Ai_q(x)$ and $Gi_q(x)$ for negative integer $q$.}
\label{f3}
\end{figure}

We can also plot the behavior of the generalized Airy and Scorer functions. For $q>0$ and an integer, the first few have the forms shown in Fig. \ref{f2}. Here again we see that both the functions are well behaved at the origin and vanish to zero as $x\rightarrow \infty$ in an oscillatory manner. Finally, for $q<0$ and an integer, the first few generalized Airy and Scorer functions have the behavior shown in Fig. \ref{f3}. They are also well behaved at the origin and vanish as $x\rightarrow \infty$ in an oscillatory manner.

\subsection{Various expansions}

First of all, we can write a formal closed form expression for the propagator ($F_{q}(x)$) in the following way. We note  from the definition \eqref{F} that
\begin{align}
F_{q}(x) & = \frac{1}{\pi}\int_{q}^{\infty} du\,e^{-i(ux+\frac{u^{3}}{3})}\notag\\
& = \frac{1}{\pi} \int_{0}^{\infty} du\,e^{-i ((u+q)x + \frac{(u+q)^{3}}{3})},\label{closedform0}
\end{align}
through a simple shift of the variable of integration. The integrand can be rearranged to give
\begin{align}
F_{q}(x) & = \frac{1}{\pi}\,e^{-i(qx+\frac{q^{3}}{3})}\,e^{(iq\frac{d^{2}}{dx^{2}}+q^{2}\frac{d}{dx})}\int_{0}^{\infty} du\,e^{-i(ux+\frac{u^{3}}{3})}\notag\\*
& =  e^{-i(qx+\frac{q^{3}}{3})}\,e^{(iq\frac{d^{2}}{dx^{2}}+q^{2}\frac{d}{dx})} (Ai(x) - iGi(x))\notag\\
& = e^{-i(qx+\frac{q^{3}}{3})}\,e^{(iq\frac{d^{2}}{dx^{2}}+q^{2}\frac{d}{dx})} F(x),\label{closedform}
\end{align}
where we have used \eqref{airy} as well as \eqref{Fzero}. Alternatively, we can also write \eqref{closedform} as
\begin{align}
F_{q}(x) & = e^{-i(qx+\frac{q^{3}}{3})}e^{iq\frac{d^{2}}{dx^{2}}} (Ai (x+q^{2}) - iGi (x+q^{2}))\notag\\*
& = e^{-i(qx+\frac{q^{3}}{3})}e^{iq\frac{d^{2}}{dx^{2}}} F(x+q^{2}),\label{closedform1}
\end{align}
where we have used the relation
\begin{equation}
e^{a\frac{d}{dx}}\, f(x) = f(x+a).
\end{equation}
Both \eqref{closedform} and \eqref{closedform1} reduce to
\begin{equation}
F_{0}(x) = F(x) = Ai(x) - i Gi(x).
\end{equation}

Using the identity
\begin{equation}
\int du\,e^{f(u)} = \frac{e^{f(u)}}{f'(u)} \sum_{n=0}^{\infty} \left(-\frac{d}{du} \frac{1}{f'(u)}\right)^{n},
\end{equation}
we note that if we identify $f(u) = -i(ux+u^{3}/3), f'(u) = -i(u^{2}+x)$, we can obtain a useful expansion of the propagator for $q^{2}+x>1$ as
\begin{align}
F_{q}(x) & = \frac{1}{\pi} \int_{q}^{\infty} du\,e^{-i(ux+\frac{u^{3}}{3})}\notag\\
& =\left.\frac{1}{\pi}\frac{e^{-i(ux+\frac{u^{3}}{3})}}{-i(u^{2}+x)} \sum_{n=0}^{\infty}\left(-\frac{d}{du} \frac{1}{-i(u^{2}+x)}\right)^{n}\right|^{\infty}_{q}\notag\\
& = \frac{e^{-i(qx+\frac{q^{3}}{3})}}{i\pi(q^{2}+x)}\left[1 + \frac{2iq}{(q^{2}+x)^{2}} + \cdots\right].
\end{align}

Similarly, when $q^{2} \ll 3|x|$, the propagator ($F_{q}(x)$) can be expanded in inverse powers of $x$ as follows. We note that we can write
\begin{align}
F_{q}(x) & = \frac{1}{\pi} \int_{q}^{\infty} du\,e^{-i(ux+\frac{u^{3}}{3})}\notag\\
& = \frac{1}{\pi}\int_{0}^{\infty} du\,e^{-i(ux+\frac{u^{3}}{3})} - \frac{1}{\pi}\int_{0}^{q} du\,e^{-i(ux+\frac{u^{3}}{3})}\notag\\
& = Ai(x)-iGi(x) - \frac{1}{i\pi x} \int_{0}^{iqx} d\bar{u}\,e^{-\bar{u}} e^{\frac{\bar{u}^{3}}{3x^{3}}},
\end{align}
where we have defined $\bar{u} = iux$. The second exponential inside the integral can be Taylor expanded leading to
\begin{equation}
F_{q}(x) = Ai(x) - iGi(x) - \frac{1}{i\pi x} \sum_{n=0}^{\infty}\frac{1}{n!}\frac{1}{(3x^{3})^{n}}\gamma(3k+1,iqx),\label{incompletegamma0}
\end{equation} 
where we have used the definition of the incomplete gamma function
\begin{equation}
\gamma (a,b) = \int_{0}^{b} dt\,e^{-t}\,t^{a-1}.\label{incompletegamma}
\end{equation}
When $q=0$, the incomplete gamma function in \eqref{incompletegamma0} vanishes, $\gamma(a,0)=0$ (as can be seen from \eqref{incompletegamma}), so that we recover the familiar result \eqref{Fzero}
\begin{equation}
F_{0}(x) = F(x) = Ai(x) - iGi(x).
\end{equation} 
We note that, for fixed values of the parameters $a_{0},a_{1},a_{2}$, all these expansions correspond to nonperturbative expansions in the sense that, for $g$ small (see \eqref{defn}), $q\sim g^{\frac{1}{3}}$ (small) whereas $x\sim g^{-\frac{1}{3}}$ (large). 

We can also obtain a perturbative expansion for the propagator ($F_{q}(x)$) as follows. We note from \eqref{closedform0} that we can write
\begin{align}
F_{q}(x) & = \frac{e^{-i(qx+\frac{q^{3}}{3})}}{\pi}\int_{0}^{\infty} du\,e^{-iu(x+q^{2}) - i(u^{2}q+\frac{u^{3}}{3})}\notag\\
& =  \frac{e^{-i(qx+\frac{q^{3}}{3})}}{\pi}\int_{0}^{\infty} du\,e^{-iu(x+q^{2})}\notag\\
&\qquad\times\sum_{k=0}^{\infty} \frac{(-i)^{k}}{k!} (u^{2}q+\frac{u^{3}}{3})^{k}\notag\\
& = \frac{e^{-i(qx+\frac{q^{3}}{3})}}{\pi}\int_{0}^{\infty} du\,e^{-iu(x+q^{2})}\notag\\
& \qquad\times\sum_{k=0}^{n}\sum_{n=0}^{k}\frac{(-i)^{k}}{n! (k-n)!} (u^{2}q)^{k-n} (\frac{u^{3}}{3})^{n}\notag\\
& = \frac{e^{-i(qx+\frac{q^{3}}{3})}}{\pi}\sum_{k=0}^{\infty}\sum_{n=0}^{k}\frac{(-i)^{k}}{n! (k-n)!}\frac{q^{k-n}}{3^{n}}\notag\\&\qquad\times\frac{(-i)^{2k+n+1}}{(x+q^{2})^{2k+n+1}}\Gamma(2k+n+1),\label{perturbative}
\end{align}
where we have used the definition of the Gamma function
\begin{equation}
\int_{0}^{\infty} dt\,e^{-\mu t}\,t^{\nu-1} = \frac{1}{\mu^{\nu}}\,\Gamma(\nu).
\end{equation}
From the definitions \eqref{defn} and \eqref{defn1}, it is easy to see that, for small $g$ and fixed values of the parameters, each term in the sum goes as $\sim (g)^{k+\frac{1}{3}}$ and thereby gives a perturbative expansion of $F_{q}(x)$.

It is interesting to note from \eqref{perturbative} that when $q=0$, namely,  $(a_{1}=0)$, the only term in the $n$ expansion \eqref{perturbative} that contributes is the term with $n=k$ and we have
\begin{align}
F_{0}(x) & = Ai(x) - iGi(x)\notag\\
& = \frac{1}{\pi}\sum_{k=0}^{\infty} \frac{(-i)^{4k+1}}{k! 3^{k}x^{3k+1}}\Gamma(3k+1)\notag\\
& = -\frac{i}{\pi x}\sum_{k=0}^{\infty} \frac{(3k)!}{k!} \frac{1}{(3x^{3})^{k}}.\label{largeexpansion}
\end{align}
As $g\rightarrow 0, x\rightarrow g^{-\frac{1}{3}}$ becomes large (see \eqref{defn1} with $q=0$). Therefore, \eqref{largeexpansion} provides a large $x$ expansion of $F(x)$ and from \eqref{largeexpansion} we can now identify
\begin{align}
Ai(x) & \xrightarrow{x\, \text{large}} 0,\notag\\
Gi(x) & \xrightarrow{x\, \text{large}} \frac{1}{\pi x}\sum_{k=0}^{\infty} \frac{(3k)!}{k!} \frac{1}{(3x^{3})^{k}}.
\end{align}
The first of these relations is known and is easily seen from Fig. \ref{f1} and the second gives a quantitative description of the decrease in the Scorer function for large value of the argument. 

\section{Conclusion}
We have studied the fermion propagator in an external time-dependent potential in $0+1$ dimension. We 
have shown that, when this potential is quadratic in time,
one can obtain a closed form result for the fermion propagator in 
terms of generalized Airy functions. These represent new generalizations of the Airy functions and reduce to the standard Airy functions when the potential is independent of a linear term in time. We have discussed various properties associated with these generalized functions.

We note here that the present analysis can also be extended to the study of the fermion propagator in an external potential at finite temperature \cite{dasfinite} where generalized Airy functions also appear. In this case, \eqref{propagatoreqn} can be solved exactly, subject to the anti-periodic boundary conditions of a finite-temperature fermion propagator \cite{das1,das2}. 
\newline

A.L.M.B. and J.F. would like to thank CNPq (Brazil) for financial support.

\appendix
\section{Solving the differential equation}

We can alternatively determine the propagator by solving the energy space differential equation in \eqref{S(p)eqn}
\begin{equation}
(p-m-gA(-i\frac{d}{dp}))S(p) = i.\label{S(p)eqn1}
\end{equation}
In this appendix we do this systematically and compare the results with the coordinate space calculations. In the zeroth order when there is no interaction ($g=0$), the free propagator
\begin{equation}
S_{0}(p) = \frac{i}{p-m},
\end{equation}
has already been discussed in \eqref{S_{0}(p)} in both the ways.

If we include only the constant term in the potential, $A(t)=a_{0}$, then \eqref{S(p)eqn1} takes the form
\begin{equation}
(p-m-ga_{0})S(p) = (p-M)S(p) = i,
\end{equation}
which determines
\begin{equation}
S(p) = \frac{i}{p-M},
\end{equation}
where we have identified, as in \eqref{M}, $M=m+ga_{0}$. This first order propagator coincides with \eqref{a_{0}}.

At the next order, $A(t)=a_{0}+a_{1}t$, the differential equation \eqref{S(p)eqn1} takes the form
\begin{align}
& (p-m-ga_{0} -ga_{1}(-i\frac{d}{dp}))S(p) = i,\notag\\
{\rm or,}\quad & (\frac{d}{dp} - \frac{i}{ga_{1}} (p-M))S(p) = \frac{1}{ga_{1}}.
\end{align}
This can be solved with an integrating factor. For example, if we define
\begin{equation}
S(p) = e^{\frac{i}{2ga_{1}} (p-M)^{2}}\, \overline{S}(p),
\end{equation}
this leads to
\begin{equation}
\frac{d\overline{S}(p)}{dp} = \frac{1}{ga_{1}}\,e^{-\frac{i}{2ga_{1}} (p-M)^{2}},
\end{equation}
which has the solution
\begin{align}
\overline{S} (p) & = C + \sqrt{\frac{2}{iga_{1}}}\int_{0}^{\sqrt{\frac{i}{2ga_{1}}} (p-M)} dp'\,e^{-(p')^{2}}\notag\\
& = C - \sqrt{\frac{\pi}{2iga_{1}}}\, \Phi (\sqrt{\frac{i}{2ga_{1}}}\, (M-p)).
\end{align}
Here $C$ is a constant and we have used the definition of the probability function or the error function as well as its antisymmetry (see \eqref{prob}) \cite{GR}
\begin{equation}
\Phi(x) = \frac{2}{\sqrt{\pi}}\int_{0}^{x} dt\, e^{t^{2}},\quad \Phi(-x) = - \Phi(x).
\end{equation}

The constant $C$ is determined from the requirement that
\begin{align}
S(p) & = e^{\frac{i}{2ga_{1}} (p-M)^{2}}\, \overline{S}(p)\notag\\
& \xrightarrow{a_{1}\rightarrow 0} \frac{i}{p-M}.
\end{align}
From the asymptotic behavior of the probability function (see \eqref{probexpansion})
\begin{equation}
\Phi(x)\xrightarrow{x\rightarrow \infty} 1 - \frac{1}{\sqrt{\pi}}\frac{e^{-x^{2}}}{x} + \cdots,
\end{equation}
we determine that 
\begin{equation}
C = \sqrt{\frac{\pi}{2iga_{1}}}.
\end{equation}
This leads to the propagator at the second order to be
\begin{equation}
S(p) = \sqrt{\frac{\pi}{2iga_{1}}}e^{\frac{i}{2ga_{1}} (p-M)^{2}} (1- \Phi(\sqrt{\frac{i}{2ga_{1}}}\, (M-p))),
\end{equation}
which coincides with \eqref{a_{1}}. 

At the next order, $A(t)=a_{0} + a_{1}t + a_{2}t^{2}$, the differential equation \eqref{S(p)eqn1} has the form
\begin{align}
& (p-m-ga_{0}-ga_{1}(-i\frac{d}{dp}) -ga_{2}(-i\frac{d}{dp})^{2})S(p) = i,\notag\\
{\rm or,}\ & (ga_{2}\frac{d^{2}}{dp^{2}} + iga_{1}\frac{d}{dp} + p-M)S(p) = i.
\end{align}
As before (see \eqref{defn}-\eqref{defn1}), we can define
\begin{equation}
q=\frac{ga_{1}}{2(ga_{2})^{\frac{2}{3}}},\quad x=-\frac{(p-M)}{(ga_{2})^{\frac{1}{3}}}-q^{2},
\end{equation}
so that the differential equation takes the form
\begin{equation}
(\frac{d^{2}}{dx^{2}} - 2iq\frac{d}{dx} - (x+q^{2})) S(x) = \frac{i}{(ga_{2})^{\frac{1}{3}}}.
\end{equation}

To solve this equation, let us define
\begin{equation}
S(x) = e^{i(qx+\frac{q^{3}}{3})}\,\overline{S}(x),
\end{equation}
so that $\overline{S}(x)$ satisfies the equation
\begin{equation}
\frac{d^{2}\overline{S}(x)}{dx^{2}} - x\overline{S}(x) = \frac{i e^{-i(qx+\frac{q^{3}}{3})}}{\pi (ga_{2})^{\frac{1}{3}}}.
\end{equation}
Comparing this with \eqref{Feqn}, we conclude that
\begin{align}
\overline{S}(x) & = \frac{\pi}{(ga_{2})^{\frac{1}{3}}}\, F_{q}(x),\notag\\
S(x) & = e^{i(qx+\frac{q^{3}}{3})} \overline{S}(x) = \frac{\pi e^{i(qx+\frac{q^{3}}{3})}}{(ga_{2})^{\frac{1}{3}}} F_{q}(x),
\end{align}
which can be compared with \eqref{prop} and \eqref{F}.


\begin{thebibliography}{10}

\bibitem{AS}  M. Abramowitz and I. A. Stegun, {\em Handbook of Mathematical Functions}, Dover Publications, New York  (1970).

\bibitem{dasquantum} A. Das and A. C. Melissinos, {\em Quantum Mechanics: A Modern Introduction}, Gordon and Breach, New York (1986).

\bibitem{bali} G. Bali, QCD forces and heavy quark bound states, Phys. Rep. {\bf 343}, 1 (2001).

\bibitem{sumino} Y. Sumino, QCD potential as a ``Coulomb-plus-linear" potential, Phys. Lett. {\bf B571}, 173 (2003); Y. Sumino, Static QCD potential at $r<\Lambda_{QCD}^{-1}$: Perturbative expansion and operator expansion, Phys. Rev. {\bf D76}, 114009 (2007).

\bibitem{drazin} P. G. Drazin and W. H. Reid, {\em Hydrodynamic Stability}, Cambridge University Press, Cambridge (1981).

\bibitem{swanson} C. A. Swanson and V. B. Headley, SIAM J. App. Mathematics {\bf 15}, 1400 (1967).

\bibitem{Chin} R. C. Y. Chin and G. W. Hedstrom, A dispersionless analysis for difference schemes: Tables of generalized Airy functions, Math. Comp. {\bf 32}, 1163 (1978).

\bibitem{Janson} S. Janson, D. E. Knuth, T. \L uczak and B. Pittel, The birth of the giant component, Random Structures Algorithms, {\bf 4}, 231 (1993).

\bibitem{Kamimoto} J. Kamimoto, On an integral of Hardy and Littlewood, Kyushu J. Mathematics {\bf 52}, 249 (1998).

\bibitem{GR} I. S. Gradshteyn and I. M. Ryzhik, {\em Tables of Integrals, Series and Products}, Academic Press, New York (1980).

\bibitem{dasfinite} A. Das, {\em Finite Temperature Field Theory}, World Scientific, Singapore (1997).

\bibitem{das1} A. Das and J. Frenkel, Finite temperature effective actions, Phys. Lett. {\bf B680}, 195 (2009).

\bibitem{das2} A. Das and J. Frenkel, Effective actions at finite temperature, Phys. Rev. {\bf D80}, 125039 (2009).


\end{thebibliography}
\end{document}